\documentstyle[12pt]{article}
\def\be{\begin{equation}}
\def\ee{\end{equation}}
\def\bea{\begin{eqnarray}}
\def\eea{\end{eqnarray}}

\begin{document}

\begin{center}
{\Large{\bf Noncommutativity and Its Freedoms Due to the PP-Wave and
Background Gauge Fields}}

\vskip .5cm
{\large Davoud Kamani}
\vskip .1cm
 {\it Institute for Studies in Theoretical Physics and
Mathematics (IPM)
\\  P.O.Box: 19395-5531, Tehran, Iran}\\
{\sl e-mail: kamani@theory.ipm.ac.ir}
\\
\end{center}

\begin{abstract}
In this article we consider an open string attached to a D$_p$-brane,
in the presence of the pp-wave and background gauge fields. The
effects of the string mass on the open string propagator, open string
metric and the noncommutativity parameter are studied. Some free
matrices appear in the propagator and in the open string variables.
The symmetries of the string propagator and consistency with the
zero mass case put some restrictions on these matrices.
\end{abstract}



\vskip .5cm
\newpage
\section{Introduction}
It is known that the plane wave metric supported by a Ramond-Ramond
5-form background \cite{1} provides examples of exactly solvable string
models \cite{2}. Many properties of string theory in such plane wave
backgrounds have been deeply investigated \cite{2,3}.
In the other side, the noncommutativity of a D-brane worldvolume has
been studied through the open strings in the presence of the background
gauge fields \cite{4}. In this article we shall study it by computing
the propagator of a massive open string.

We shall consider an open string in the presence of both pp-wave and
background gauge fields.
This produces a generalized noncommutativity on the
D-brane worldvolume that the
open string is attached on it. This generalization comes from the
propagator equation which is a second order differential equation with
source term. The solution of this equation has expression in
terms of the (modified) Bessel functions. Beside the background fields,
the string mass also appears in the noncommutativity
parameter, in the open string metric and in the string propagator.

In addition, some free matrices also arise in these variables.
Symmetries of the open string propagator and consistency with the known
cases put some restrictions on these matrices. For example, imposing
the Hermitian condition on the string propagator leads to a restriction
on these matrices and also gives a
noncommutativity that for the zero gauge fields is not zero.
In the motion of a D$_p$-brane along itself such free matrices also arise
in the open string variables and in the propagator \cite{5}.

Since the bosonic string theory is sufficient to discuss the
noncommutativity of the spacetime, we shall consider only the bosonic
open strings.

The paper is organized as follows. In section 2, from the open string
action we obtain the open string variables and the equation of the
string propagator. In section 3, the propagator equation will be solved.
In section 4, the freedoms of the matrices that
appear in the open string variables will be restricted.
\section{Open string variables}
In the background gauge fields and pp-wave the action of open string
can be written as in the following
\bea
S&=&\frac{1}{4\pi \alpha'}\int d\tau \int_0^{2\pi\alpha' p^+}d\sigma
\bigg{(}g_{IJ}(\partial_\tau X^I \partial_\tau X^J + \partial_\sigma X^I
\partial_\sigma X^J + \mu^2 X^IX^J ) \bigg{)}
\nonumber\\
&~&-\frac{i}{2}\int d\tau ({\cal{F}}_{\alpha \beta} X^\alpha
\partial_\tau X^\beta )_{\sigma_0}\;,
\eea
where $g_{IJ}$ is a constant metric and
${\cal{F}}_{\alpha \beta}$ is total field strength, i.e.,
${\cal{F}}_{\alpha \beta}=\partial_\alpha A_\beta-\partial_\beta A_\alpha
+B_{\alpha \beta}$. The field $A_\alpha$ is a $U(1)$ gauge field with
constant field strength. Therefore, in the action (1)
it has been expressed in the form
$A_\alpha = -\frac{1}{2}F_{\alpha \beta} X^\beta$.
Furthermore, we assumed that the NS$\otimes$NS field $B_{\alpha \beta}$
is constant and the
metric of the worldsheet is Euclidean. The parameter $\sigma_0=0$
indicates the end of the open string on the brane. Let us
consider the decomposition $\{X^I\}=\{X^\alpha\} \bigcup \{X^i\}$.
The set $\{X^\alpha\}$ shows the brane directions and $\{X^i\}$ are
coordinates perpendicular to the brane.

The variation of the action (1) gives the equation of motion and
the boundary conditions of the open string coordinates
\bea
(\partial^2_\tau +\partial^2_\sigma-\mu^2) X^I(\sigma , \tau)=0\;,
\eea
\bea
(\partial_\sigma X^\alpha + 2\pi i\alpha'{\cal{F}}^\alpha_{\;\;\;\beta}
\partial_\tau X^\beta)_{\sigma_0}=0\;,
\eea
\bea
(\delta X^i)_{\sigma_0}=0\;.
\eea
As we see the mass ``$\mu$'' does not affect the boundary conditions.

According to the equations (2) and (3) the open string propagator
has the equations
\bea
\bigg{(}\partial {\bar \partial}-\frac{\mu^2}{4}\bigg{)}
{\cal{G}}^{\alpha \beta}(z, z') =
- \frac{\pi}{2}\alpha' G^{\alpha \beta} \delta^{(2)}(z-z')\;,
\eea
\bea
\bigg{(} (\partial -{\bar \partial}){\cal{G}}^{\alpha \beta}(z,z') + 
2\pi \alpha'{\cal{F}}^\alpha _{\;\;\;\gamma}(\partial+{\bar \partial})
{\cal{G}}^{\gamma \beta}(z,z') \bigg{)}_{z={\bar z}} = 0\;,
\eea
where $z=\tau+i\sigma$ and $G^{\alpha \beta}$ is the open string metric.
The solution of the propagator equations can be written in the form
\bea
{\cal{G}}^{\alpha \beta}(z,z') &=& -\alpha' \bigg{[} \frac{1}{2} 
Q^{\alpha \beta} \ln(z-z')+ \frac{1}{2} (Q^T)^{\alpha \beta}
\ln({\bar z}-{\bar z'})
\nonumber\\
&~& +\bigg{(}-\frac{1}{2}Q^{\alpha \beta} + G^{\alpha \beta} +
\frac{\theta^{\alpha \beta}}{2\pi \alpha'}\bigg{)} \ln(z-{\bar z'})
\nonumber\\
&~& +\bigg{(}-\frac{1}{2}(Q^T)^{\alpha \beta} + G^{\alpha \beta} 
-\frac{\theta^{\alpha \beta}}{2\pi \alpha'}\bigg{)} \ln({\bar z}- z')
-\frac{i}{2\alpha'} D^{\alpha \beta} \bigg{]}\;,
\eea
where $\theta^{\alpha \beta}$ is the noncommutativity parameter.
For the next purposes let the matrix $D^{\alpha \beta}$
be anti-Hermitian
\bea
D^\dagger =-D\;.
\eea
In addition, we assume that $D^{\alpha \beta}$ is constant but
the matrix $Q^{\alpha \beta}$ depends on the variables
$z, {\bar z}, z'$ and ${\bar z'}$.

Combining the boundary equation (6) and the propagator (7), we obtain
the open string metric and the noncommutativity parameter
\bea
G^{\alpha \beta} = \bigg{(}(g+2 \pi \alpha' {\cal{F}})^{-1}g 
(g-2 \pi \alpha' {\cal{F}})^{-1} R \bigg{)}^{\alpha \beta}\;,
\eea
\bea
G_{\alpha \beta} = \bigg{(}R^{-1}(g-2 \pi \alpha' {\cal{F}}) g^{-1}
(g+2 \pi \alpha' {\cal{F}})\bigg{)}_{\alpha \beta}\;,
\eea
\bea
\theta^{\alpha \beta} = -(2\pi \alpha')^2 \bigg{(}(g+2 \pi \alpha'
{\cal{F}})^{-1}{\cal{F}}(g-2 \pi \alpha' {\cal{F}})^{-1} R 
\bigg{)}^{\alpha \beta}\;.
\eea
In these variables the matrix $R$ has the definition
\bea
R \equiv \frac{1}{2}\bigg{(} (g+2\pi\alpha'{\cal{F}})Q
+(g-2\pi\alpha'{\cal{F}})Q^T \bigg{)}_{z={\bar z}}\;.
\eea
The open string variables $G$ and $\theta$
depend on the mass $\mu$ through the matrix
$Q$. According to the propagator (7) for the zero mass $\mu$ we should
have $Q=g^{-1}$. In this case there is $R={\bf 1}$, as expected.

The matrix $Q$ should satisfy the boundary condition
\bea
\bigg{(} (g+2\pi\alpha'{\cal{F}})\partial(Q-Q^T)
-(g-2\pi\alpha'{\cal{F}}){\bar \partial}(Q-Q^T)
\bigg{)}_{z={\bar z}}=0\;.
\eea
In fact, the equation (6) leads to the relations (9)-(11) and the
equation (13). Without lose of generality, we assume that the matrix
$Q$ for all coordinates $z$ and ${\bar z}$ to be symmetric, i.e.,
\bea
Q^T = Q \;.
\eea
Therefore, the equation (13) becomes trivial, and the matrix
$R$ takes the form $R=gQ|_{z={\bar z}}$.
\section{Solution of the propagator equation}
According to the propagator (7), the equation (5) reduces to the
matrix equation
\bea
\Omega + \Omega^\dagger = 0,
\eea
where the matrix $\Omega$ is defined by
\bea
\Omega(z,{\bar z})=\bigg{(}\partial {\bar \partial}
-\frac{\mu^2}{4}\bigg{)}(\lambda Q)-
\frac{\mu^2}{4}\bigg{(}-\frac{i}{2\alpha'}D+
2(g+2\pi \alpha'{\cal{F}})^{-1}R\;\ln(z-{\bar a})\bigg{)}\;.
\eea
In this definition the relation
$\bigg{(}G+\frac{\theta}{2\pi \alpha'}\bigg{)}^{\alpha\beta}
=\bigg{(}(g+2\pi \alpha'{\cal{F}})^{-1}R\bigg{)}^{\alpha\beta}$
has been used. The function $\lambda(z)$ is
\bea
\lambda(z) = \ln \bigg{(} \frac{z-a}{z-{\bar a}}\bigg{)}\;.
\eea
Since in the equation (15) $z'$ and ${\bar z'}$ are not variables,
we demonstrated them as the parameters $a$ and ${\bar a}$, respectively.

The general solution of the equation (15) can be written as in the
following
\bea
\Omega^{\alpha\beta} = \frac{\mu^2}{4}E^{\alpha\beta}_\mu
(z,{\bar z})\;,
\eea
where $E^{\alpha\beta}_\mu (z,{\bar z})$ is any anti-Hermitian matrix
that depends on the mass ``$\mu$''.
The symmetric condition (14) and reduction to the massless case 
will restrict this matrix.
The restrictions on $E_\mu^{\alpha\beta}$ will be discussed in the next
section. In fact, ${\cal{G}}^{\alpha\beta}$, $G^{\alpha\beta}$ and
$\theta^{\alpha\beta}$ are depended on the mass $\mu$
through the equation (18).

Note that the equation (18) also can be obtained as the equation of motion
from the following action
\bea
S_h=\frac{1}{4\pi \alpha'}\int d^2 z ( \partial h_{\alpha\beta}
{\bar \partial}h^{\alpha\beta}+\frac{\mu^2}{4}h_{\alpha\beta}
h^{\alpha\beta}+h_{\alpha\beta}J^{\alpha\beta})\;,
\eea
where $h^{\alpha\beta}$ and $J^{\alpha\beta}$ are
\bea
&~& h^{\alpha\beta}= \lambda Q^{\alpha\beta}\;,
\nonumber\\
&~& J^{\alpha\beta}=
\frac{\mu^2}{2}\bigg{[}-\frac{i}{2\alpha'}D^{\alpha \beta}+
2\bigg{(}(g+2\pi \alpha'{\cal{F}})^{-1}R\bigg{)}^{\alpha\beta}
\ln(z-{\bar a})+E_\mu^{\alpha\beta}(z, {\bar z})\bigg{]}\;.
\eea
Also we have $h_{\alpha\beta}= g_{\alpha\gamma} g_{\beta\lambda}
h^{\gamma\lambda}$.
In this action the elements of the matrix $Q^{\alpha\beta}$ are
degrees of freedom. Therefore, in the equation (18) the matrix
$J^{\alpha\beta}$ can be interpreted as a source for
the field $h^{\alpha\beta}$.

Define the worldsheet coordinates $\zeta$ and ${\bar \zeta}$ as
\bea
\zeta = z-{\bar a}\;\;\;,\;\;\;{\bar \zeta}={\bar z}-a\;.
\eea
In terms of these variables the equation (18) takes the form
\bea
\bigg{(}\frac{\partial}{\partial\zeta}\frac{\partial}
{\partial{\bar \zeta}}-\frac{\mu^2}{4}\bigg{)}
h^{\alpha \beta}(\zeta , {\bar \zeta}) =
\frac{\mu^2}{4}\bigg{[}-\frac{i}{2\alpha'}D^{\alpha \beta}+
2\bigg{(}(g+2\pi \alpha'{\cal{F}})^{-1}R\bigg{)}^{\alpha\beta}\ln\zeta
+E_\mu^{\alpha\beta}(\zeta, {\bar \zeta})\bigg{]}\;.
\eea
In other words, in terms of the polar forms of $\zeta$ and ${\bar \zeta}$,
i.e.,
\bea
\zeta=re^{i\phi}\;\;\;,\;\;\;{\bar \zeta}=re^{-i\phi}\;,
\eea
the equation (22) changes to
\bea
\bigg{(}\partial^2_r + \frac{1}{r}\partial_r+\frac{1}{r^2}
\partial^2_\phi -\mu^2 \bigg{)} h^{\alpha \beta}(r , \phi) =
\mu^2\bigg{[}-\frac{i}{2\alpha'}D^{\alpha \beta}
\nonumber\\
+2\bigg{(}(g+2\pi \alpha'{\cal{F}})^{-1}R\bigg{)}^{\alpha\beta}(\ln r
+i\phi)+E_\mu^{\alpha\beta}(r, \phi)\bigg{]}\;.
\eea

The solution of the equation (24) can be written as
\bea
h^{\alpha \beta}(r , \phi)=h_0^{\alpha \beta}(r , \phi)+
h_1^{\alpha \beta}(r , \phi)\;,
\eea
where $h_0^{\alpha \beta}(r , \phi)$ is the solution of the
homogeneous equation
\bea
\bigg{(}\partial^2_r + \frac{1}{r}\partial_r+\frac{1}{r^2}
\partial^2_\phi -\mu^2 \bigg{)} h_0^{\alpha \beta}(r , \phi) =0\;.
\eea
The inhomogeneous part
$h_1^{\alpha \beta}(r , \phi)$ completely depends on the right
hand side of the equation (24). On the other hand, it has
the integral form
\bea
h_1^{\alpha \beta}(r , \phi) = \frac{1}{2\pi} \int d^2 {\bf r'}
G({\bf r} , {\bf r'}) J_1^{\alpha\beta}({\bf r'})\;,
\eea
where ${\bf r} = (r , \phi)$ denotes a vector. In this integral
the vector ${\bf r'}$ sweeps all the plane.
Also $G({\bf r} , {\bf r'})$ is Green's function which satisfies
the equation
\bea
\bigg{(}\partial^2_r + \frac{1}{r}\partial_r+\frac{1}{r^2}
\partial^2_\phi -\mu^2 \bigg{)} G({\bf r} , {\bf r'}) =
2\pi \delta^{(2)}({\bf r}-{\bf r'})\;.
\eea
The source $J_1^{\alpha \beta}(r , \phi)$ is defined by the right
hand side of the equation (24). In other words, we have
$J_1^{\alpha\beta} = 2J^{\alpha\beta}$. From now on, in
$J_1^{\alpha\beta}$ we apply the approximation $R \approx {\bf 1}$. 

The Green's function equation (28) has the solution
\bea
G({\bf r} , {\bf r'})= -K_0(\mu |{\bf r}- {\bf r'}|)\;,
\eea
where $K_0(x)$ is a modified Bessel function. This function has the
integral representation
\bea
K_0(x) = \int_0^\infty d t \cos(x\sinh t)\;\;\;\;, \;\;\;\;{\rm for}
\;\;\;\; x>0\;.
\eea
Note that for $\mu|{\bf r}-{\bf r'}| \rightarrow \infty$, the
Green's function (29) goes to zero.

The equation (26) can be solved by the method of separation of the
variables. The final result is
\bea
h_0^{\alpha \beta}(r , \phi)= \sum_{n=0}^\infty \bigg{[}\bigg{(}
A_n^{\alpha \beta}\cos(n\phi) + B_n^{\alpha \beta}\sin(n\phi)
\bigg{)}I_n(\mu r) \bigg{]}\;.
\eea
The index ``$n$'' comes from the separation of the variables
$r$ and $\phi$. The functions $\{I_n(x)\}$ are modified Bessel
functions. They have the series form 
\bea
I_n(x) = \sum_{l=0}^\infty \frac{1}{l!(l+n)!}\bigg{(}
\frac{x}{2}\bigg{)}^{2l+n}\;.
\eea
Since for the integer ``$n$'' we have $I_n(x)=I_{-n}(x)$,
there is no need of adding $I_{-n}(x)$ to the solution (31).
Also $\{A^{\alpha\beta}_n\;,\;B^{\alpha\beta}_n\}$
are arbitrary constant matrices. In the next section, some restrictions
will be put on them. For the massless case the solution (31) reduces to
$h^{\alpha\beta}_0=A^{\alpha\beta}_0$.

The homogeneous solution (31) describes the modification of the open
string variables due to the string mass. While the inhomogeneous
solution (27) gives the effects of the combination of the string mass
and gauge fields to these variables.

The (modified) Bessel functions $K_0(x)$ and
$I_n(x)$ for the small and large $x$ have opposite behaviors.
For $x \rightarrow 0$ we have $K_0(x)\rightarrow +\infty$, while
$I_n(x)\rightarrow 0$ (for $n \neq 0$) and $I_0(x)\rightarrow 1$.
Also for $x \rightarrow +\infty$ there are $K_0(x)\rightarrow 0$ and
$I_n(x)\rightarrow +\infty$.
\section{Some restrictions on the matrices $A_n^{\alpha\beta}$,
$B_n^{\alpha\beta}$ and $E_\mu^{\alpha\beta}$}

In the limit $\mu \rightarrow 0$ we should have
$Q^{\alpha\beta}(z,{\bar z})=g^{\alpha\beta}$. According to the
propagator (7) this should hold for all $z$
and ${\bar z}$. It is not restricted to the $z={\bar z}$ case that appeared
in the matrix $R^{\alpha\beta}$.
This imposes the following restriction on the matrices
$A_n^{\alpha\beta}$, $B_n^{\alpha\beta}$ and $E_\mu^{\alpha\beta}(r,\phi)$,
\bea
&~&\int d^2 {\bf r'} \bigg{(} \mu^2 K_0(\mu|{\bf r}-{\bf r'}|)
E_\mu^{\alpha\beta}({\bf r'}) \bigg{)}_{\mu=0}= 2\pi(A_0^{\alpha\beta}
-\lambda g^{\alpha\beta})
\nonumber\\
&~&-\int d^2 {\bf r'} \bigg{[}\bigg{(} \mu^2 K_0(\mu|{\bf r}
-{\bf r'}|) \bigg{)}_{\mu=0}
\bigg{(}-\frac{i}{2\alpha'}D^{\alpha \beta}+
2\bigg{(}(g+2\pi \alpha'{\cal{F}})^{-1} \bigg{)}^{\alpha\beta}(\ln r'
+i\phi')\bigg{)}\bigg{]}\;.
\eea
Note that $|{\bf r}-{\bf r'}|$ can be infinite. This implies that for
$\mu \rightarrow 0$ the argument $\mu|{\bf r}-{\bf r'}|$ in
general is not small. Therefore, in this equation and also in the
next equations, we cannot use the expansion form
of the function $K_0(\mu|{\bf r}-{\bf r'}|)$ for the small argument.

Another restriction on these matrices comes from the symmetry of
the matrix $Q^{\alpha\beta}$. On the other hand, we have
\bea
&~&\mu^2\int d^2 {\bf r'} \bigg{(} K_0(\mu|{\bf r}-{\bf r'}|)
[E_\mu({\bf r'})-E^T_\mu({\bf r'})]^{\alpha\beta}\bigg{)}=
2\pi\bigg{(}h_0^{\alpha\beta}(r , \phi) - {(h^T_0)}^{\alpha\beta}
(r , \phi)\bigg{)}
\nonumber\\
&~&-\mu^2\int d^2 {\bf r'} \bigg{[}K_0(\mu|{\bf r}
-{\bf r'}|) \bigg{(}\frac{2\theta_0^{\alpha\beta}}{\pi \alpha'}(\ln r'
+i\phi')+\frac{i}{2\alpha'}(D^T-D)^{\alpha \beta}\bigg{)}\bigg{]}\;,
\eea
where $\theta_0^{\alpha\beta}$ is the noncommutativity parameter
corresponding to the massless string. From this integral equation
we obtain
\bea
&~&\mu^2 K_0(\mu|{\bf r}-{\bf r'}|)\bigg{(}
E_\mu({\bf r})-E^T_\mu({\bf r}) 
+\frac{2\theta_0}{\pi \alpha'}(\ln r
+i\phi)+\frac{i}{2\alpha'}(D^T-D)\bigg{)}^{\alpha \beta}
\nonumber\\
&~& =2\pi\bigg{(}h_0(r , \phi) - h^T_0(r , \phi)
\bigg{)}^{\alpha\beta}\delta^{(2)}({\bf r}-{\bf r'}) +
\nabla \cdot {\bf W}^{\alpha \beta}({\bf r},{\bf r'}),
\eea
where the components of the vector
${\bf W}({\bf r},{\bf r'})$ are antisymmetric matrices.
This vector, up to the following equation, is arbitrary
\bea
\int d^2 {\bf r}\nabla \cdot {\bf W}({\bf r},{\bf r'})=0\;.
\eea
Therefore, the freedom of $E_\mu({\bf r})$ is saved by the vector
${\bf W}({\bf r},{\bf r'})$. When the vector ${\bf r'}$
approaches to ${\bf r}$, both sides of the equation (35) go to
infinity. The equations (35) and (36) are consistent with (33).
That is, subtract the transpose of the equation (33) from itself.
The resulted equation is satisfied by the equations (35) and (36).

Since the matrix $R^{\alpha\beta}$ should be
independent of the worldsheet coordinate $\tau$,
we obtain another restriction on the free matrices. Note that we have
$R=gQ|_{\sigma=0}$. For $\sigma=0$ the polar coordinates are
\bea
&~&r(\tau) = \sqrt{(\tau-a)(\tau-{\bar a})}\;,
\nonumber\\
&~&\phi(\tau) = \frac{i}{2}\ln \bigg{(}\frac{\tau-a}{\tau-{\bar a}}\bigg{)}
=\frac{i}{2}\lambda(\tau)\;.
\eea
These equations show the parametric form of a curve in the
$r-\phi$ plane. In other words, $r$ versus $\phi$ is
\bea
r (\phi) = \frac{b}{\sin \phi}\;,
\eea
where $b= \frac{a-{\bar a}}{2i}$ is a real number.
In fact, motion of the end of the open string on the brane draws a curve.
Mapping of this curve in the $r-\phi$ plane leads to the equation (38).
This equation gives
\bea
Q^{\alpha\beta}(\phi) = \frac{i}{2\phi}\bigg{(}h_0^{\alpha\beta}
[r(\phi) , \phi ]+h_1^{\alpha\beta}[r(\phi) , \phi ]\bigg{)}\;.
\eea
Since $Q^{\alpha\beta}(\phi)$ should be independent of the
coordinate $\phi$, we obtain the condition
\bea
\int d^2 {\bf r'} \bigg{(} J^{\alpha\beta}({\bf r'})
(1-\phi \frac{d}{d \phi})
K_0[\mu|{\bf r}(\phi)-{\bf r'}|] \bigg{)}= \pi
(1-\phi \frac{d}{d \phi})
h_0^{\alpha\beta}[r(\phi) , \phi ]\;.
\eea
This integral equation implies that
\bea
J^{\alpha\beta}({\bf r})(1-\phi' \frac{d}{d \phi'})
K_0[\mu|{\bf r'}(\phi')-{\bf r}|]= \pi(1-\phi \frac{d}{d \phi})
h_0^{\alpha\beta}[r(\phi) , \phi ]
\;\delta^{(2)}({\bf r}-{\bf r'}) +
\nabla \cdot {\bf U}^{\alpha\beta}({\bf r},{\bf r'})\;.
\eea
The vector ${\bf U}({\bf r},{\bf r'})$ should obey the equation (36).
Subtract the transpose of the equation (41) from itself. Now eliminate
$E_\mu({\bf r})-E^T_\mu({\bf r})$ between the resulted equation
and the equation (35). Therefore, the matrices
${\bf U}({\bf r},{\bf r'})$ and ${\bf W}({\bf r},{\bf r'})$
are related to each other.

In fact, the equations (33), (35) and (41) are restrictions on
the free matrices. Up to these equations, these matrices remain arbitrary.

{\bf The Hermitian propagator}

In the propagator (7) if we use $Q^\dagger$ instead of
$Q^T$, the matrix ${\cal{G}}^{\alpha\beta}$ will be Hermitian.
This change also will appear in the equations (12)-(14).
Therefore, the matrix $Q$ should be Hermitian. This condition
restricts the free matrices by the equation
\bea
&~&\mu^2\int d^2 {\bf r'} \bigg{(} K_0(\mu|{\bf r}-{\bf r'}|)
E_\mu^{\alpha\beta}({\bf r'}) \bigg{)}= 2\pi
\frac{{\bar \lambda}h_0^{\alpha\beta}(r , \phi)
-\lambda (h_0^\dagger)^{\alpha\beta}(r , \phi)}
{\lambda + {\bar \lambda}}
\nonumber\\
&~&-\mu^2\int d^2 {\bf r'} \bigg{[} K_0(\mu|{\bf r}-{\bf r'}|)
\bigg{(} 2i G_1^{\alpha\beta} \phi' + \frac{\theta_1^{\alpha\beta}}
{\pi \alpha'}\ln r' + \frac{i}{2\alpha'}\frac{\lambda-{\bar \lambda}}
{\lambda + {\bar \lambda}}D^{\alpha\beta}\bigg{)}\bigg{]}\;.
\eea
The matrices $G_1^{\alpha\beta}$ and $\theta_1^{\alpha\beta}$
are the following combinations of
$G_0^{\alpha\beta}$ and $\theta_0^{\alpha\beta}$,
\bea
&~&G_1^{\alpha\beta} = \bigg{(}G_0-\frac{\lambda-{\bar \lambda}}
{\lambda + {\bar \lambda}}\frac{\theta_0}{2\pi \alpha'}
\bigg{)}^{\alpha\beta}\;,
\nonumber\\
&~&\theta_1^{\alpha\beta} = \bigg{(}\theta_0-2\pi\alpha'
\frac{\lambda-{\bar \lambda}}{\lambda + {\bar \lambda}}G_0
\bigg{)}^{\alpha\beta}\;,
\eea
where $G_0^{\alpha\beta}$ denotes the open string metric for the
massless case.
These imply that $G_1$ is Hermitian and $\theta_1$ is anti-Hermitian.
The matrices $G_1$ and $\theta_1$ may be interpreted as combined open
string metric and combined noncommutativity parameter.
In the absence of the gauge fields we have
$G_0^{\alpha\beta}=g^{\alpha\beta}$ and $\theta_0^{\alpha\beta}=0$,
while $\theta_1^{\alpha\beta}$ is not zero. Therefore, effectively
there is a noncommutativity.

The integral equation (42) gives the explicit form of the matrix
$E_\mu({\bf r})$,
\bea
&~&\mu^2 K_0(\mu|{\bf r}-{\bf r'}|)
\bigg{(}E_\mu({\bf r})
+2i G_1\phi + \frac{\theta_1}{\pi \alpha'}\ln r +
\frac{i}{2\alpha'}\frac{\lambda'-{\bar \lambda'}}
{\lambda' + {\bar \lambda'}}D\bigg{)}
\nonumber\\
&~&= 2\pi \frac{{\bar \lambda}h_0({\bf r})
-\lambda h_0^\dagger({\bf r})}
{\lambda + {\bar \lambda}}\;\delta^{(2)}({\bf r}-{\bf r'})+
\nabla \cdot {\bf V}({\bf r},{\bf r'})\;,
\eea
where $\lambda' = \lambda (r' , \phi')$. The vector
${\bf V}({\bf r},{\bf r'})$, that its components are anti-Hermitian
matrices, also should satisfy the equation (36).
\section{Conclusions}
Propagation of an open string which its ends are on a brane, in the
presence of the pp-wave and background gauge fields, generalizes the
noncommutativity of the brane worldvolume. The same generalization also
occurs in the open string metric and in the propagator.
This generalization has two parts.
One part completely depends on the string mass. The other part is due to
the string mass and the background gauge fields.

As we saw the propagator equation also can be obtained from an action
with the source term.
Solving the equation of motion of the string propagator, we obtained some
free matrices in ${\cal{G}}^{\alpha\beta}$, $G^{\alpha\beta}$
and $\theta^{\alpha\beta}$. The symmetries of the string propagator
and matching it with the zero mass case put some
restrictions on these matrices.
Up to some equations, these matrices remain arbitrary.
In other words, for the given background fields $g_{\alpha\beta}$,
$B_{\alpha\beta}$, $A_{\alpha}$ and $\mu$,
there is a family of the noncommutativities. Each element
of this family is distinguished by its corresponding matrices
$\{A_n^{\alpha\beta}, B_n^{\alpha\beta}, E_\mu^{\alpha\beta}\}$.

Imposing the Hermitian condition on the string propagator, we obtained
a modified open string metric and noncommutativity parameter.
In this case for the vanishing gauge fields there is a noncommutativity,
proportional to the closed string metric.


\begin{thebibliography}{99} 
\bibitem{1} 
M. Blau, J. Figueroa-O' Farrill, C. Hull and G. Papadopoulos, JHEP
{\bf 0201}(2002)047, hep-th/0110242.
\bibitem{2}
R.R. Metsaev, Nucl. Phys. {\bf B625}(2002)70, hep-th/011044; R.R. Metsaev
and A.A. Tseytlin, Phys. Rev. {\bf D65}(2002)126004, hep-th/0202109.
\bibitem{3}
D. Berenstein, J. Maldacena and H. Nastase, JHEP {\bf 0204}(2002)013,
hep-th/0202021;
D. Amati and C. Klimcik, Phys. Lett. {\bf B210}(1988)92; G.T. Horowitz
and A.R. Steif, Phys. Rev. Lett. {\bf 64}(1990)260; H.J. de Vega and
N. Sanchez, Phys. Rev. Lett. {\bf 65}(1990)1517, Phys. Lett. {\bf B244}
(1990)215; O. Jofre and C. Nunez, Phys. Rev. {\bf D50}(1994)5232,
hep-th/9311187; C.R. Nappi and E. Witten, Phys. Rev. Lett. {\bf 71}
(1993)3751, hep-th/9310112; E. Kiritsis and C. Kounnas, Phys. Lett.
{\bf B320}(1994)264, hep-th/9310202; C. Klimcik and A.A. Tseytlin,
Phys. Lett. {\bf B323}(1994)305, hep-th/9311012; K. Sfetsos and A.A.
Tseytlin, Nucl. Phys. {\bf B427}(1994)245, hep-th/9404063; J.G. Russo
and A.A. Tseytlin, Nucl. Phys. {\bf B448}(1995)293, hep-th/9411099,
JHEP {\bf 0204}(2002)021, hep-th/0202179;
G.T. Horowitz and A.R. Steif, Phys. Rev. {\bf D42}(1990)1950; G.T.
Horowitz and A.A. Tseytlin, Phys. Rev. {\bf D51}(1995)2896, hep-th/9409021;
R. Brooks, Mod. Phys. Lett. {\bf A6}(1991)841; H.J. de Vega and
N. Sanchez, Phys. Rev. {\bf D45}(1992)2783; H.J. de Vega, M. Ramon Medrano
and N. Sanchez, Class. Quant. Grav. {\bf 10}(1993)2007; G. Papadopoulos,
J.G. Russo and A.A. Tseytlin, hep-th/0211289; J.G. Russo and A.A. Tseytlin,
JHEP {\bf 0209}(2002)035, hep-th/0208114; C.S. Chu and P.M. Ho, Nucl.
Phys. {\bf B636}(2002)141, hep-th/0203186;
D. Kamani, Phys. Lett. {\bf B564}(2003)123, hep-th/0304236;
``{\it Strings in the Quantized pp-wave Backgrounds from Membrane}'',
hep-th/0301003.
\bibitem{4}
N. Seiberg and E. Witten, JHEP {\bf9909}(1999)032, hep-th/9908142;
A. Connes, M.R. Douglas and A. Schwarz, JHEP {\bf 9802}(1998)003,
hep-th/9711162; M.R. Douglas and C. Hull, JHEP {\bf 9802}(1998)008,
hep-th/9711165;  
C.S. Chu and P.M. Ho, Nucl. Phys. {\bf B550}(1999)151,
hep-th/9812219; Nucl. Phys. {\bf B568}(2000)447, hep-th/9906192;
A. Fayyazuddin and M. Zabzine, Phys. Rev. {\bf D62}(2000)046004,
hep-th/9911018; P.M. Ho and Y.T. Yeh, Phys. Rev. Lett. {\bf 85}(2000)5523,
hep-th/0005159;
Y.E. Cheung and M. Krog, Nucl. Phys. {\bf B528}(1998)185, hep-th/9803031;
D. Bigatti and L. Susskind, Phys. Rev. {\bf D62}(2000)066004, 
hep-th/9908056; A. Schwarz, Nucl. Phys.
{\bf B534}(1998)720, hep-th/9805034;
F. Ardalan, H. Arfaei and M.M. Sheikh-Jabbari, JHEP {\bf 9902}(1999)016,
hep-th/9810072; Nucl. Phys. {\bf B576}(2000)578, hep-th/9906161;
V. Schomerus, JHEP {\bf 9906}(1999)030, hep-th/9903205;
D. Kamani, Phys. Lett. {\bf B548}(2002)231, hep-th/0210253.
\bibitem{5}
D. Kamani, Europhys. Lett. {\bf 57}(2002)672, hep-th/0112153.

\end{thebibliography}
\end{document}